\documentclass[a4paper,fleqn]{cas-dc}
\usepackage[numbers]{natbib}
\usepackage{tikz}

\begin{document}
\let\WriteBookmarks\relax
\def\floatpagepagefraction{1}
\def\textpagefraction{.001}
\let\printorcid\relax
\shorttitle{Diverse cooperation tendencies}

\shortauthors{Linya Huang et~al.}

\title [mode = title]{Evolution of cooperation with the diversity of cooperation tendencies}                

% \author[number] 中 number 对应 \address 编号，通讯作者直接在 \author[number]{name} 花括号内 name 后面加上 \corref{cor1}，并使用 \cortext[cor1]{Corresponding author} 进行通信作者的说明。
\author[1]{Linya Huang}

\author[1,2]{Wenchen Han}\corref{cor1}

\affiliation[1]{College of Physics and Electronic Engineering, Sichuan Normal University, Chengdu, 610101, People's Republic of China.}

\affiliation[2]{Lanzhou Center for Theoretical Physics, Key Laboratory of Theoretical Physics of Gansu Province, Lanzhou University, Lanzhou, Gansu 730000, People's Republic of China.}

\cortext[cor1]{Email: wchan@sicnu.edu.cn}

\begin{abstract}
The complete cooperation and the complete defection are two typical strategies considered in evolutionary games in many previous works. 
However, in real life, strategies of individuals are full of variety rather than only two complete ones. 
In  this work, the diversity of strategies is introduced into the weak prisoners' dilemma game, which is measured by the diversity of the cooperation tendency. 
A higher diversity means more cooperation tendencies are provided. 
The complete cooperation strategy is the full cooperation tendency and the complete defection strategy is without any cooperation tendency. Agents with other cooperation tendencies behave as partial cooperators and as partial defectors simultaneously. 
The numerical simulation shows that increasing the diversity of the cooperation tendency promotes the cooperation level, not only the number of cooperators but also the average tendency over the whole population, until the diversity reaches its saturated value. 
Furthermore, our work points out maintaining cooperation is based on the cooperation efficiency approximating to the reward of cooperators and that the cooperation efficiency oscillates and quickly decreases to zero when cooperator clusters cannot resist the invasion of defectors.
When the effect of the noise for the Femi update mechanism is considered, a higher diversity of strategies not only improves the cooperation level of the whole population but also supports the survival of more rational agents.  

\end{abstract}

% \includegraphics{figs/grabs.pdf}

%\item Research highlights item 1
%\item Research highlights item 2
%\item Research highlights item 3

% Keywords
\begin{keywords}
Cooperation tendency\sep Cooperation efficiency \sep Cooperation level \sep prisoners' dilemma
\end{keywords}
\maketitle
\section{Introduction}

As an important subject studies the decision-making and the cooperation, the game theory profoundly reveals the dynamic behaviors among individuals under different environments \cite{Fudenberg et al.1991,Myerson 1991}. Its significance extends beyond economics to fields such as social sciences and ecology \cite{Zagare 1984,Kim 2014,Barron 2024}. 
As a classic theoretical framework, the game theory is extensively applied to explain the paradox between the Darwinian principle and the universality of the cooperation, and to investigate the efficient mechanisms to promote the cooperation level \cite{Smith 1982,Colman 2013}.

Typical game models commonly studied in literature include the two-player prisoner's dilemma (PD) \cite{Rapoport 1965,Doebeli et al.2005} and the multiplayer public goods game (PGG) \cite{Szolnoki er al.2009}. 
In the former, a pair of strategies is involved and the stable equilibrium is mutual defection, resulting in the extinction of cooperators. In the latter, defectors afford no cost while cooperators bear the cooperation cost, leading to a tragedy of commons \cite{Hardin 1968,Macy 2002}. 
Although the cooperation is generally considered difficult to maintain, it is commonly observed in reality. To explore the intricate relationship between the individual rationality and the group profit, researchers have identified specific mechanisms that promote cooperation through the evolutionary game theory \cite{Nowak 2006}. 

Early studies classified these mechanisms into five main categories, that is, the kin selection \cite{Hamilton 1964,Gardner et al.2011,Wild et al.2023}, the direct reciprocity \cite{Trivers 1971,Sigmund 2010,Tkadlec et al.2023}, the indirect reciprocity \cite{Nowak 1998,Leimar 2001,Nowak 2005,Zonca et al.2021}, the network reciprocity \cite{Nowak 1992,Szolnoki 2008,Stojkoski et al.2019}, and the group selection \cite{Boyd et al.2002,Traulsen et al.2006,Civilini et al.2021}. In addition, other systemic mechanisms have emerged, including the reputation \cite{Nowak 1998,Fu 2008,Gross 2019,Shen 2022}, rewards and punishments \cite{Szolnoki 2010,Shen 2023,Helbing 2010,Sun 2023}, and the heterogeneity of agents \cite{Szolnoki 2007,Perc 2008,Bi.2023,Pan.2023}. In traditional games, there is usually only the complete cooperation strategy and the complete defection strategy to represent two opposing behaviors. However, the world is not black and white, ambiguous ideas and neutral behaviors exist. As early as 2007, Szolnoki introduced the concept of heterogeneity, and since then, more and more researchers have employed various methods to incorporate heterogeneity and explore its effects. For example, Yuan \emph{et al.} \cite{Yuan et al.2014} introduced the investment heterogeneity by relating individual investment to the local cooperation level and found the supercritical relation promotes the cooperation the most. 
In the structure of scale-free networks, Cao \emph{et al.} \cite{Cao et al.2010} introduced the investment heterogeneity by connecting individual investment to the proportion of individual connectivity and the cooperation is remarkably promoted when agents with larger degrees contribute less.
These two works map individual heterogeneity to different investments, while Yan \emph{et al.} \cite{Bi et al.2023} proposed a heterogeneous reputation evolution mechanism, which makes it easier to promote cooperation by updating strategies based on the relative reputation adjustment. 

Due to the setting of the PGG, each participant can invest freely in the group. It is natural to consider investment heterogeneity in the PGG model, but the heterogeneity in PD about the strategy is difficult to be introduced. 
Chen \emph{et al.} \cite{Chen et al.2022} defined a quasi-cooperative strategy, which permits agents being incompletely altruistic. The strategy of an agent is not confined to the complete cooperation strategy and the complete defection strategy. 
Lee \emph{et al.} \cite{Lee et al.2023} used a quasi-cooperative strategy with agents applying the myopic best response rule and found the improvement of the cooperation level. 
Pan \emph{et al.} \cite{Pan et al.2022} introduced a quasi-defection strategy with a cooperative tendency to promote group profit while preserving their interests as much as possible, which can effectively promote cooperation in highly competitive systems. 
Song \emph{et al.} \cite{Song et al.2023} studied the model with the introduction of dual attribute agents playing a snowdrift dilemma into the traditional prisoners' dilemma and found the cooperation can be promoted. 

Above works introduced a third strategy into the prisoners' dilemma, which shows the features of both the cooperation and the defection strategy but also some differences. The third strategy also requires a revised version of the payoff matrix rather than following the original one. 
Agents play a traditional prisoners' dilemma when only the full cooperation tendency, i.e. the complete cooperation, is provided. 
Based on these, this work introduces the diversity of the strategy by allowing agents adopt different cooperation tendencies into a weak prisoners' dilemma. 
Agents holding the $x$ value of the cooperation tendency will behave as a cooperator with $x$ proportion and as a defector with $1-x$ proportion simultaneously. From this point of view, there is no need to revise the payoff matrix. 
More tendency levels provide more cooperation tendencies, which permit more choices for the value of the cooperation tendency. 
It is found that the cooperation can be improved when more cooperation tendencies are provided and that there exists a saturated cooperation tendencies at a certain defection temptation. It is also found that agents should be more rational when more cooperation tendencies are provided.

In the following sections, we present a detailed model of the prisoner's dilemma based on the cooperation tendency in Section \ref{sec2}. Subsequently, we present the simulation results in Section \ref{sec3}, accompanied by in-depth discussions. The final section summarizes the main findings of the study and provides conclusions.

\section{Model}\label{sec2}

In this study, we focus on the evolution of a population of agents playing the weak prisoner's dilemma (wPD) \cite{Nowak 1992,Szabó et al.1998} located on an $L\times L$ Lattice with periodic boundary conditions, where agents interact with their adjacent agents with a module $L$. For example, an agent sitting on $(1, L)$ is connected with agents on $(2,L)$, $(L,L)$, $(1,1)$, and $(1,L-1)$. In this lattice, the neighbor of agent $i$ is denoted as the set $N_i$ and the number of neighbour set elements is the degree of each node $k=4$.
In the wPD, each agent has two strategies, the cooperation and the defection. Each agent will get reward $1$ when two cooperators play the game, agents get punishment $0$ when two agents are defectors, and the defector can get the defection temptation $b\in [1,2)$ but the cooperator get the sucker $0$ when two agents hold different strategies. 
In reality, agents are not entirely reliant on a single strategy but hold a tendency towards the cooperation strategy maintaining a degree of uncertainty or hesitancy. 
%Such as heterogeneous investments on public goods games \cite{Yuan et al.2014,Cao et al.2010,Bi et al.2023}, quasi-cooperators \cite{Lee et al.2023} and quasi-defectors \cite{Pan et al.2022} in PDs. 
Here, we introduce the diversity of the cooperation tendency $m$ into the wPD to provide different levels of cooperation tendencies, where cooperation tendencies of agents are denoted as $S=\{0,1/m,2/m,\dots,1\}$. That is, agent $i$ will can be seen as $s_i$ part of a cooperator and $1-s_i$ part of a defector simultaneously. 
When $m=1$, it degenerates to a common case in traditional evolutionary games. There are only two extreme tendencies about the cooperation, i.e., $s=0$ for the complete defection and $s=1$ for the complete cooperation. While $m\ge 2$, agents can be partial cooperative together with partial defective.  
When agent $i$ with a tendency $s_i$ and agent $j$ with $s_j$ play the wPD together, the payoff of agent $i$ is $\pi_{i,j}=s_i\cdot s_j+(1-s_i)\cdot s_j\cdot b$ and the payoff of agent $j$ is $\pi_{j,i}=s_j\cdot s_i+(1-s_j)\cdot s_i\cdot b$. Generally, $\pi_{i,j}=\pi_{j,i}$ if and only if $s_i=s_j$. The total payoff of agent $i$ on the lattice is 
\begin{equation}    \label{eq1}
\pi_i=\sum_{j\in N_i}\pi_{i,j}. 
\end{equation}
Agent $i$ learns the tendency of agent $j$, randomly chosen from agent $i$'s neighbours, with the probability 
\begin{equation}
 W(s_j\leftarrow s_i)=\frac{1}{1+\exp[(\pi_i-\pi_j)/\kappa]},
\end{equation}
where $\kappa$ is the noise \cite{Szabó et al.1998,Szabó et al.2005}, characterizing the rationality of agents. $\kappa\to 0$ means agents are extremely rational, where a small payoff difference will lead agents to update their strategy, and $\kappa\to\infty$ means agents are extremely irrational, where agents learns others strategy with a probability $0.5$ independent of their payoff difference. $\kappa=0.1$ is fixed where agents are rather rational unless specified. 

With the introduction of cooperation tendency, the cooperation level can be measured not only by the fraction of cooperators
\begin{equation}
    f_C=\frac{|\{j|s_j>0\}|}{N},
\end{equation}
where $|\{x\}|$ represents the cardinality of the set $\{x\}$, but also by the average cooperation tendency
\begin{equation}
    f_S=\frac{\sum_j s_j}{N}.
\end{equation}
$f_C$ denotes the ratio of cooperative agents to the total number of agents, while $f_S$ indicates the average cooperation tendency within the entire population.
Since $f_C$ treats all cooperative agents with different cooperation tendencies as equal and $f_S$ accounts for their specific tendencies, it follows that $f_C\ge f_S$, with $f_C=f_S$ if and only if all cooperative agents exhibit complete cooperation tendency $s=1$. Thus, $f_S/f_C\in (0,1]$ characterizes the average cooperation tendency among cooperative agents.
Additionally, the cooperation efficiency $e_C$ is defined as the average payoff per cooperation tendency across the entire population, given by
\begin{equation}
    e_C=\frac{\langle\pi\rangle}{4f_S},
\end{equation}
where $\langle\pi\rangle=\frac{\sum_j \pi_j}{N}$ denotes the average payoff throughout the population, and $\frac{\langle\pi\rangle}{4}$ represents the average payoff for each game. Obviously, $e_C=1$ for all cooperators and $e_C=0$ for all defectors.

In this study, $N=10000$ agents are positioned on a $100\times 100$ lattice. Once all agents have gathered their payoffs, they simultaneously update their cooperation tendencies. Presented results are averages over $1000$ Monte Carlo time steps (MCTS) in the steady state, after $50000$ transition time steps, averaged over $50$ random initial conditions unless otherwise noted.

\section{Simulation and Analysis}\label{sec3}

\begin{figure}% 
\centerline{\includegraphics[width=2.6\linewidth]{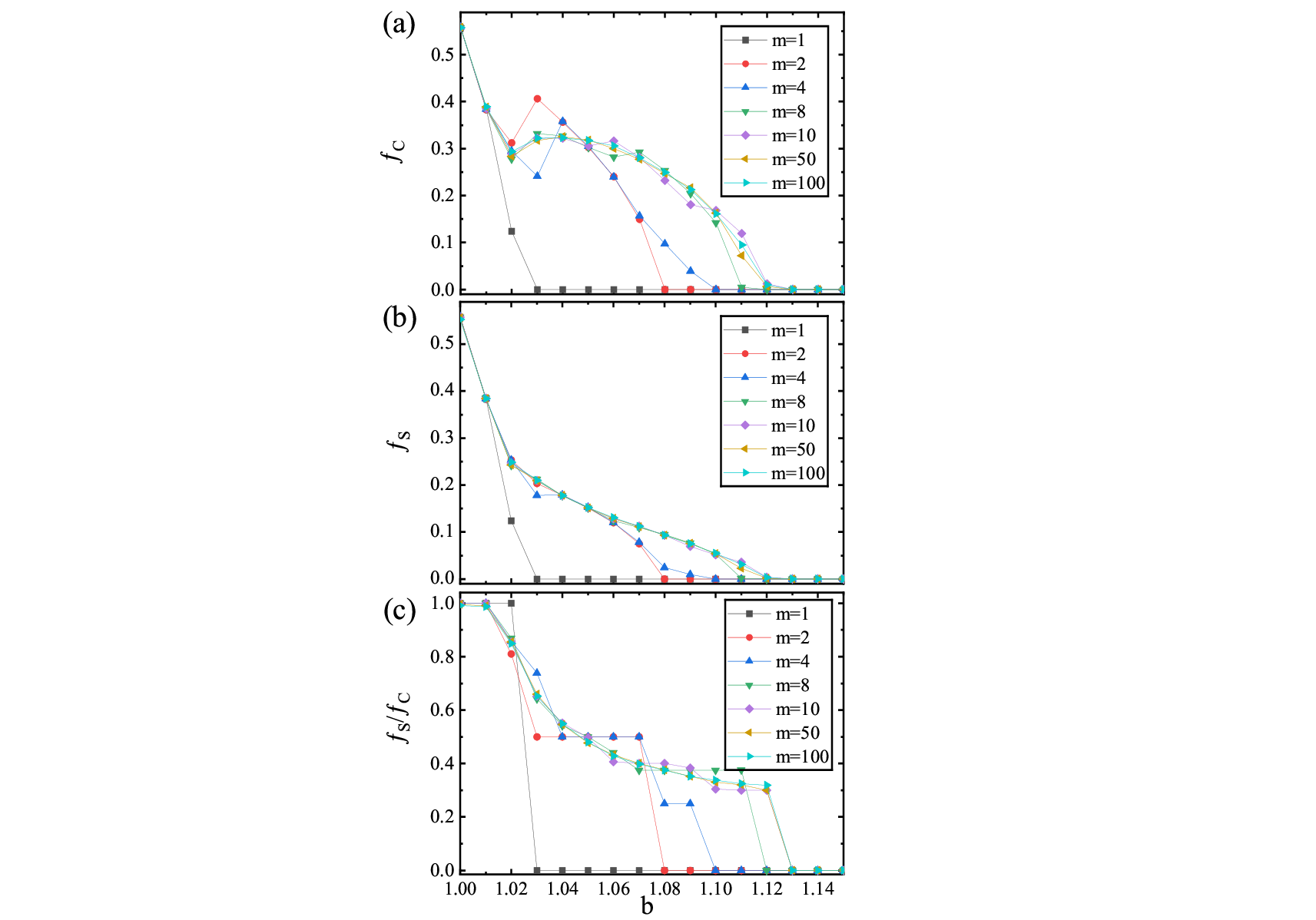}}
\caption{The diagram of the cooperation against the defection temptation $b$ with different tendency parameters $m$. 
(a) The fraction of cooperators $f_C$ and (b) the average cooperation tendency $f_S$. (c) The average tendency of cooperators $f_S/f_C$.}\label{fig1}
\end{figure}

\begin{figure*}[h]%
\centering
\includegraphics[width=0.9\textwidth]{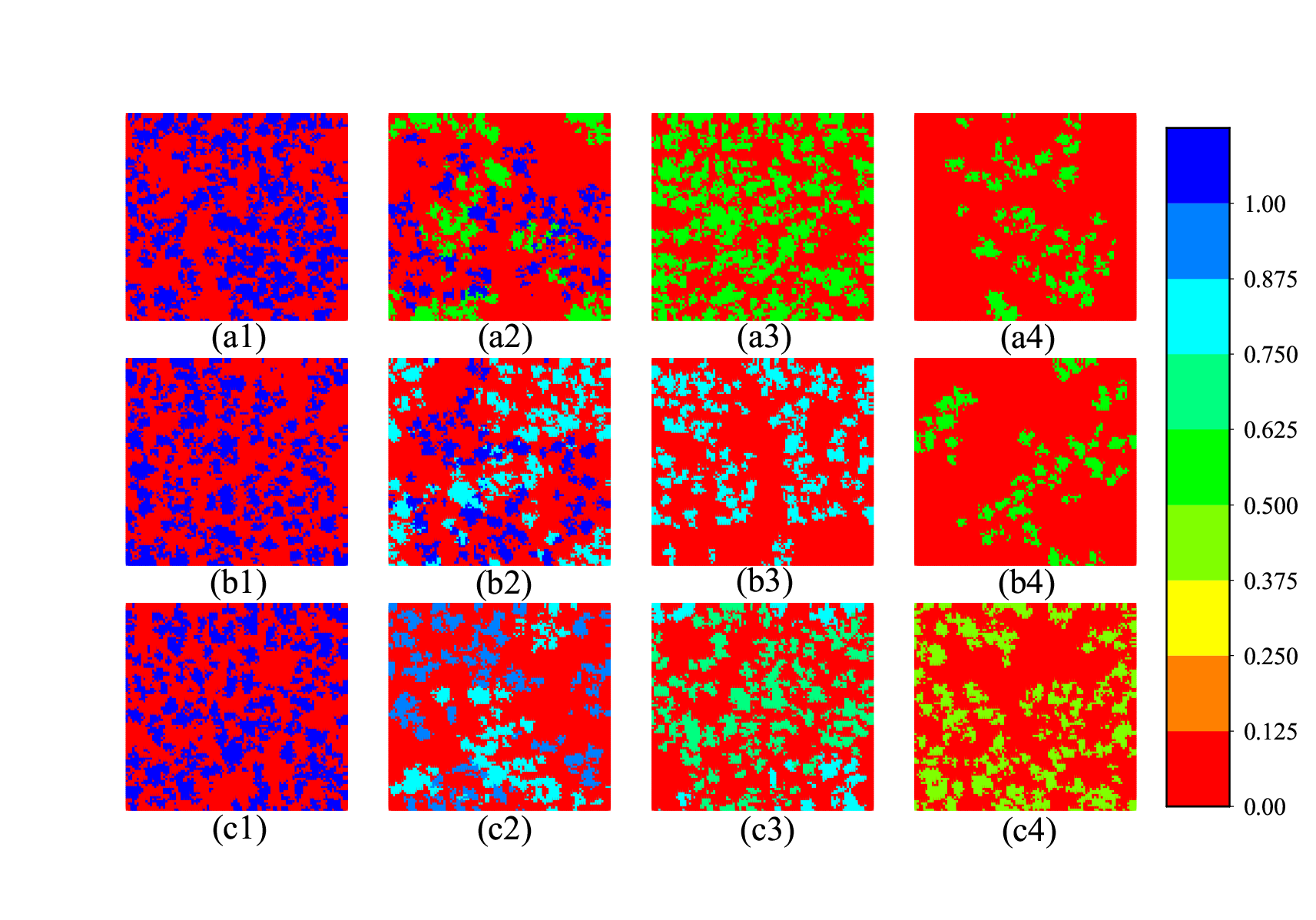}
\caption{Snapshots of agents' cooperation tendencies in the steady state with different parameters at $10000$ MCTS. 
The cooperation tendency diversity $m=2$ in (a), $m=4$ in (b), and $m=8$ in (c). 
Different columns for defection temptations $b=1.01$, $b=1.02$, $b=1.03$, and $b=1.07$.}\label{fig2}
\end{figure*}

The fraction of cooperators $f_C$ is used as the cooperation level in many works concerned with the evolutionary game theory \cite{Zhu et al.2019,He JL.2022}. The effects of $b$ on $f_C$ is shown in Fig.~\ref{fig1} (a). $m=1$ is the case for agents with the complete defection or the complete cooperation strategy and all agents will prefer the defection when the temptation is a little bit higher $b>1.02$. 
When agents have more cooperation tendency diversities $m\ge 2$, $f_C$ shows a non-monotone trend and $f_C$ finally decreases to $0$ until the temptation is large $b\approx 1.13$. It shows clearly that more tendency diversities (a larger $m$) support the the population in reaching a higher cooperation level overall. However, when the temptation is just a little higher about $b=1.02$ for $m>1$, it seems that $m=2$ supports the cooperation most but $m=4$ are the worst at $b=1.03$ and that $m=2$ and $m=4$ supports the cooperation more than others at $b=1.04$. 
This is quite out of intuition that how is the phenomenon triggered, which cannot be understood just from the fraction of cooperators $f_C$ and indicates that details of the mechanism for this phenomenon are hidden. 
Then the average cooperation tendency $f_S$ is investigated and the relation upon the temptation $b$ is shown in Fig.~\ref{fig1} (b). Contrary to the $f_C$, $f_S$ shows a monotone trend where a higher temptation $b$ makes agents prefer the defection more. 
A higher cooperation tendency diversity $m$ gives a higher opportunity of cooperators' survival unless $m$ is saturated, which indicates that there exist the highest level of cooperation even for the support of the network reciprocity in spatial lattices \cite{Nowak 1992,Szolnoki 2008,Stojkoski et al.2019}, where clustered cooperators can survive from the invasion of defectors. 
Figure~\ref{fig1} (c) shows the average cooperation tendency over cooperators $f_S/f_C$ against the defection temptation $b$. 
It shows clearly that when agents only have complete strategies with $m=1$ cooperators just has two values, i.e., $1$ for the complete cooperation and $0$ for no cooperators survival. 
When $m=2$, there are three values of $f_S/f_C$, i.e., $1$ for the complete cooperation, $0$ for the complete defection, and $0.5$ for the partial cooperation. These three values are shown in Fig.~\ref{fig1} (c) and they have formed three main stages. 
However, there is another value of $f_S/f_C\approx 0.8$ at $b=1.02$ for $m=2$ indicating the coexistence of agents with $s=1$ and $s=0.5$. 
Due to coexistence of cooperative agents with $s=1$ and $s=0.5$, at $b=1.02$ partial cooperators $s=0.5$ survive and the average cooperation tendency $f_S$ can be promoted towards the saturated one, compared with a lower level of $f_S$ thanks to no partial cooperation choice provided for $m=1$.  
Besides increasing the temptation $b\in [1.03,1.08]$ hinders the cooperation, fewer cooperators with $s=0.5$ survive, where the population with $m=1$ only prefers the defection. This perfectly shows the cooperation level of the population is supported by partial cooperators. 
When $m=4$, cooperative agents can also choose $s=0.25$ or $s=0.75$ as their strategy, it shows cooperators prefer $s=0.75$ rather than $s=0.5$ at $b=1.03$ when the choice is provided, which means $s=0.75$ is more close to the saturated average cooperation tendency of cooperators. 
While the highest cooperation tendency of the whole population is defined, a higher cooperation tendency only asks for fewer cooperators, which is the reason why $f_C$ with $m=2$ is higher than that with $m=4$ in Fig.~\ref{fig1} (a). 
It is the same with the population provided $m=8$ at $b=1.07$. A higher cooperation tendency diversity allows agents choose the cooperation with tendency $s=0.375$, which supports agents to form the cooperation tendency much closer to the highest one, permitting more cooperators.
However, as the highest cooperation tendency is defined, the average tendency of cooperators $f_S/f_C$ merge into one when the cooperation tendency diversity $m$ increases further. 

Snapshots of cooperation tendencies in Fig.~\ref{fig2} show the detail that which cooperation tendencies are cooperators' favorite one or two when they are provided more $m\ge2$. 
When the temptation $b=1.01$ is small, cooperators prefer the tendency $s=1$ even when more tendencies are provided shown in Fig.~\ref{fig2} (a1), (b1), and (c1). 
When the temptation $b=1.02$, cooperators only have to choose the mix of $s=1$ and $s=0.5$ when only two cooperation tendencies are provided $m=2$ shown in Fig.~\ref{fig2} (a2), while cooperators prefer the mix of $s=1$ and $s=0.75$ for $m=4$ in Fig.~\ref{fig2} (b2), but cooperators would like $s=0.75$ and $s=0.875$ for $m=8$ in Fig.~\ref{fig2} (c2) at the given time step. 
Only cooperators with $s=0.875$ for $m=8$ will survive when the evolution time is longer.  
This indicates that cooperators would prefer the mix of cooperation tendencies or a pure cooperation tendency no higher than the saturated cooperation tendency rather than the mix of all cooperation tendencies. 
As for $b=1.03$, the third column of Fig.~\ref{fig2} shows it much clearly that a little bit higher temptation cannot support the choice of the complete cooperation $s=1$ but the partial one $s=0.5$ with $m=2$ in Fig.~\ref{fig2} (a3), which is much lower than the saturated average cooperation of cooperators and permits more cooperators to survive. 
When $m=4$ in Fig.~\ref{fig2} (b3), cooperators prefer $s=0.75$ close to the saturated average cooperation tendency $f_S$ and a higher cooperation tendency makes a lower fraction of cooperators $f_C$ due to the limitation of $f_S$. More cooperation tendency are provided by $m=8$ in Fig.~\ref{fig2} (c3), $s=0.625$ are more preferable to $s=0.75$, and more cooperators survive. 
It is the same as $b=1.07$ for in Fig.~\ref{fig2} (a4), (b4), and (c4) but with a different saturated $f_S$.

\begin{figure}[h]% 
\centerline{\includegraphics[width=1.7\linewidth]{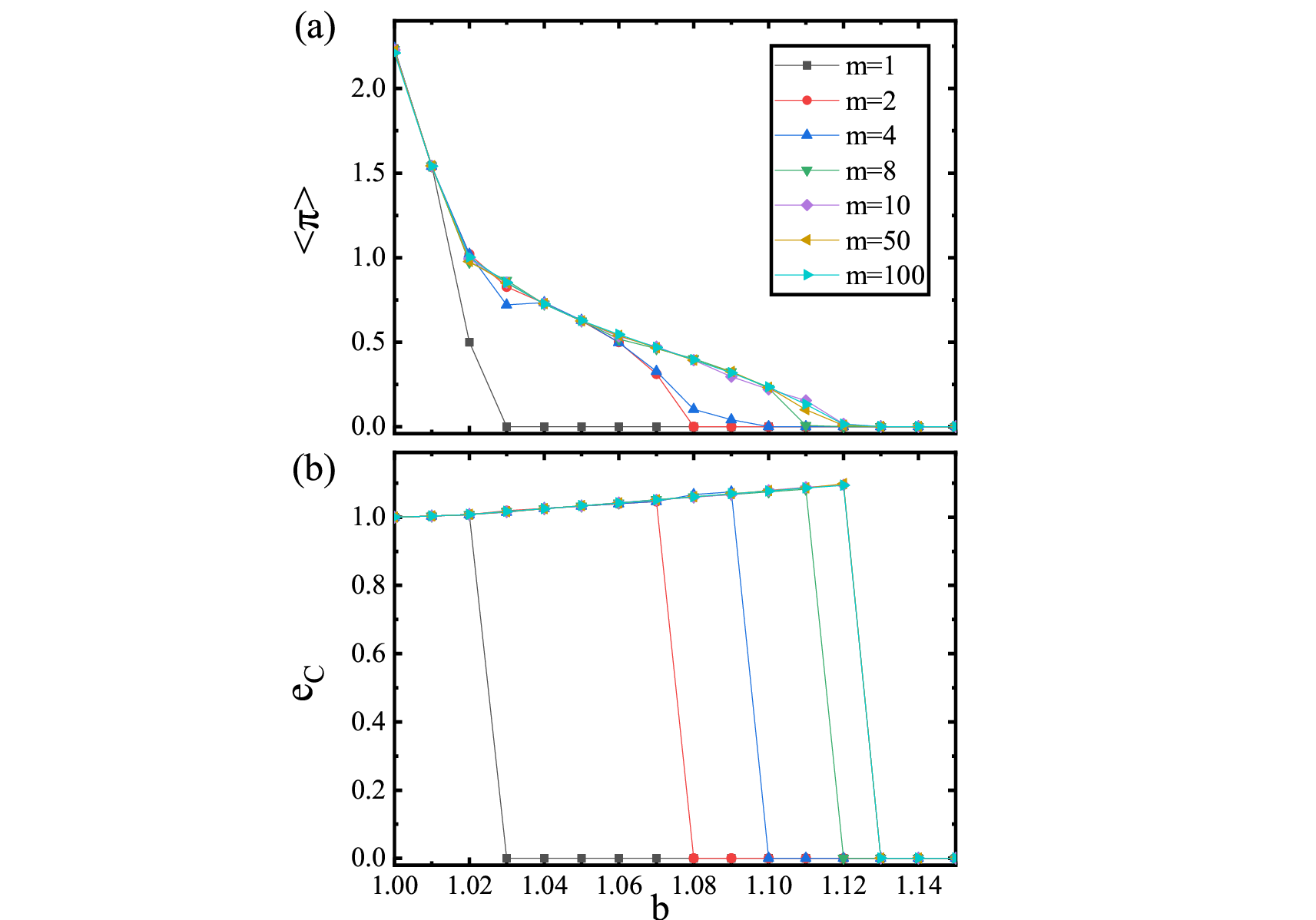}}
\caption{ (a) The average payoff over the whole population $\langle\pi\rangle$ and (b) the cooperation efficiency (the average payoff per cooperation tendency) $e_C=\frac{\langle\pi\rangle}{4f_S}$ against the defection temptation $b$ with different cooperation tendency levels $m$.}\label{fig3}
\end{figure}

\begin{table}[]
    \centering
\begin{tabular}{|c|c|c|}
\hline Case & Configuration & $e_C$  \\ \hline
A&\begin{tikzpicture}
    \draw[help lines, color=gray!60] (-1,-1) grid (1,1);
    \fill (0,0) circle(0.1);  \node at (0,0) [right] {C};
    \draw(0,1) circle (0.1);  \draw (0,-1) circle(0.1); \draw (1,-1) circle(0.1); \draw (1,1) circle(0.1);
    \draw (1,0) circle(0.1); \draw (-1,0) circle(0.1); \draw (-1,-1) circle(0.1); \draw (-1,1) circle(0.1);
\end{tikzpicture}& $b>1$     \\  \hline
B&\begin{tikzpicture}
    \draw[help lines, color=gray!60] (-1,-1) grid (2,1);
    \fill (0,0) circle(0.1);  \node at (0,0) [right] {C};   \fill (1,0) circle(0.1);  \node at (1,0) [right] {C};
    \draw(-1,1) circle (0.1);  \draw (-1,0) circle(0.1); \draw (-1,-1) circle(0.1); \draw (0,1) circle(0.1);
    \draw (2,1) circle(0.1); \draw (2,0) circle(0.1); \draw (2,-1) circle(0.1); \draw (0,-1) circle(0.1);
    \draw (1,1) circle(0.1); \draw (1,-1) circle(0.1); 
\end{tikzpicture}& $\frac{1+3b}{4}>1$     \\ \hline
C&\begin{tikzpicture}
    \draw[help lines, color=gray!60] (-1,-1) grid (2,2);
    \fill (0,0) circle(0.1);  \node at (0,0) [right] {C};   \fill (1,0) circle(0.1);  \node at (1,0) [right] {C};
    \fill (0,1) circle(0.1);  \node at (0,1) [right] {C};   \fill (1,1) circle(0.1);  \node at (1,1) [right] {C};
    \draw(-1,1) circle (0.1);  \draw (-1,0) circle(0.1); \draw (-1,-1) circle(0.1); \draw (-1,2) circle(0.1);
    \draw (2,1) circle(0.1); \draw (2,0) circle(0.1); \draw (2,-1) circle(0.1); \draw (2,2) circle(0.1);
    \draw (1,2) circle(0.1); \draw (1,-1) circle(0.1); \draw (0,2) circle(0.1); \draw (0,-1) circle(0.1); 
\end{tikzpicture}& $\frac{1+b}{2}>1$     \\ \hline
\end{tabular}
    \caption{Typical configuration cases for the cooperation efficiency $e_C$. Full nodes for cooperators and empty nodes for defectors. Case A for a cooperator surrounded by defectors, Case B for two clustered cooperators surrounded by defectors, and Case C for four clustered cooperators surrounded by defectors.}
    \label{table1}
\end{table}

%However, it is still quite strange why the average cooperation tendency $f_S$ with $m=2$ and $m=4$ is much lower than the optimum one when the temptation $b$ becomes larger shown in Fig.~\ref{fig1} (b) for $b\in[1.08,1.12)$. 
Figure~\ref{fig3} (a) shows the trend of the average payoff over the whole population $\langle\pi\rangle$ against the defection temptation and that the trend of the average payoff $\langle\pi\rangle$ is exactly the same as the trend of the average cooperation tendency $f_S$. 
Remembering the rule of the wPD, positive payoffs are all contributed by cooperators and defectors contribute nothing. Thus it is quite intuitive that the trend of the average cooperation tendency $f_S$ is maintained by cooperators and the average payoff is the support of the cooperation. 
This observation leads us to calculate the average payoff per cooperation tendency $\frac{\langle\pi\rangle}{4f_S}$, which is called the efficiency of the cooperation $e_C$ and the result is shown in Fig.~\ref{fig3} (b) with different cooperation tendency diversity $m$.
Obviously, the cooperation efficiency is $1$ for two cooperators, $0$ for two defectors, and $b/2$ for the case a cooperator encounter a defector. As $b\in[1,2]$ is the discussion range in wPD, the cooperation efficiency is at most $1$ for the pairwise interaction. 
The cooperation efficiency $e_C$ around cooperators will be larger than full cooperators as shown in Table~\ref{table1}, due to the defection temptation $b$ is higher than the cooperation reward $1$. 
This means a cooperator surrounded by more defectors on average permits a higher cooperation efficiency $e_C$, but a higher $e_C$ also leads to an unstable configuration. Cooperators will turn to be defectors in unstable configurations until the cooperation efficiency $e_C$ approaches $1$, where cooperators in clusters can support each other causing the network reciprocity. 
This rule is robust independent of the diversity of the cooperation tendency and give rise to the phenomenon that $e_C$ with different $m$ collapse onto a same line. 
Furthermore, in the final steady state the coexistence of cooperator clusters and defectors leads to cooperators at the cluster boundaries contributing to a higher $e_C$ as the temptation $b$ increases. 

\begin{figure}[h]%
\centerline{\includegraphics[width=1.2\linewidth]{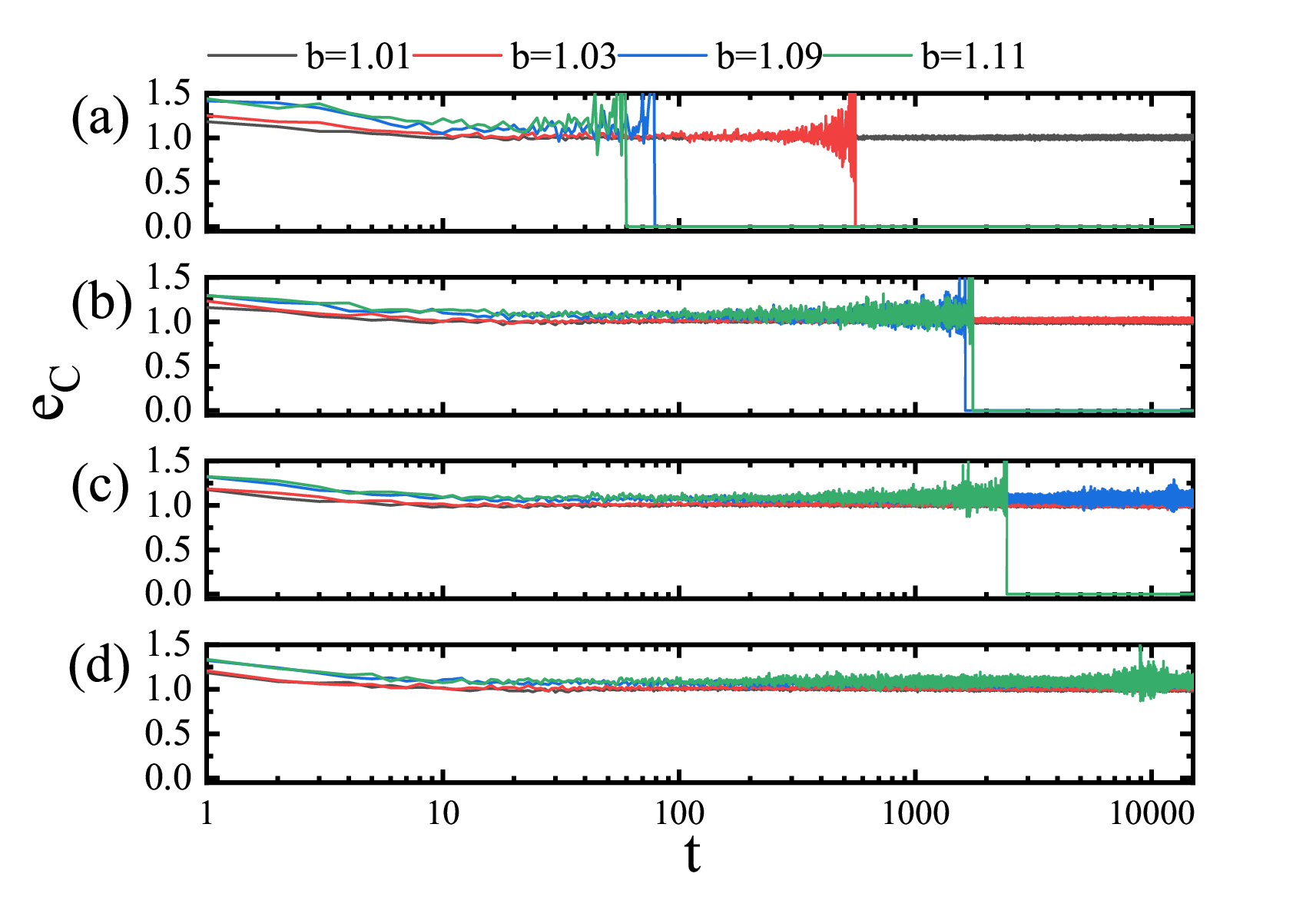}}
\caption{The evolution of the cooperation efficiency $e_C$ (the average payoff per cooperation tendency $\frac{\langle\pi\rangle}{4f_S}$) with different cooperation tendency levels $m$. The cooperation tendency level $m=1$ in (a), $m=2$ in (b), $m=4$ in (c), and $m=8$ in (d).}\label{fig4}
\end{figure}

However, there still exists an question why the cooperation efficiency $e_C$, the payoff contributed by the cooperation tendency on average $\frac{\langle\pi\rangle}{4f_S}$, with different cooperation tendency levels $m$ dropping to $0$ at different temptations $b$. 
The evolution of $e_C$ can explain a lot, shown in Fig.~\ref{fig4}.  
At the beginning, the initial condition has not reached the steady state, where cooperators suffer the 
invasion of the defection and the efficiency of the cooperation $e_C$ is a relatively high level. Along with the evolution, the reorganization makes cooperators clustered and $e_C$ decreases approaching to $1$, which is the natural equilibrium point of pair cooperators. 
The case with $m=1$ is the traditional wPD in Fig.~\ref{fig4} (a). 
If clustered cooperators can resist the invasion of the defection, the cooperation efficiency stays around the balance point $1$, as shown in Fig.~\ref{fig4} (a) with $b=1.01$. While clustered cooperators are invaded by defectors, $e_C$ abruptly oscillates to a large value and quickly jumps to $0$, where no cooperators survives, as shown in Fig.~\ref{fig4} (a) with $b=1.03, 1.09, 1.11$. 
When $m=2$ in Fig.~\ref{fig4} (b), cooperators can choose the partial cooperation $s=0.5$. More cooperation tendencies introduced into the system alleviate the oscillation of $e_C$, indicating that the half cooperation tendency may act as a medium to slow down the payoff variation and the cooperation efficiency $e_C$ to reach a steady state. 
Obviously, the time for $e_C$ reaching $0$ is greatly postponed at $b=1.09, 1.11$ with $m=2$ in Fig.~\ref{fig4} (b), compared with $m=1$ in Fig.~\ref{fig4} (a), and $e_C$ stays around $1$ with cooperator clusters remaining with $b=1.03$. 
When provided with more cooperation tendency levels $m$, the population can remain some cooperators even when they suffer a relative high defection temptation for $b=1.09$ with $m=4$ in Fig.~\ref{fig4} (c) and $b=1.11$ with $m=8$ in  Fig.~\ref{fig4} (d). 
Figure~\ref{fig4} shows that the introduction of the cooperation diversity can promote the cooperation due to providing more cooperation tendencies to slow down the variation of the cooperation efficiency. 

\begin{figure}[h]%
\centerline{\includegraphics[width=1.1\linewidth]{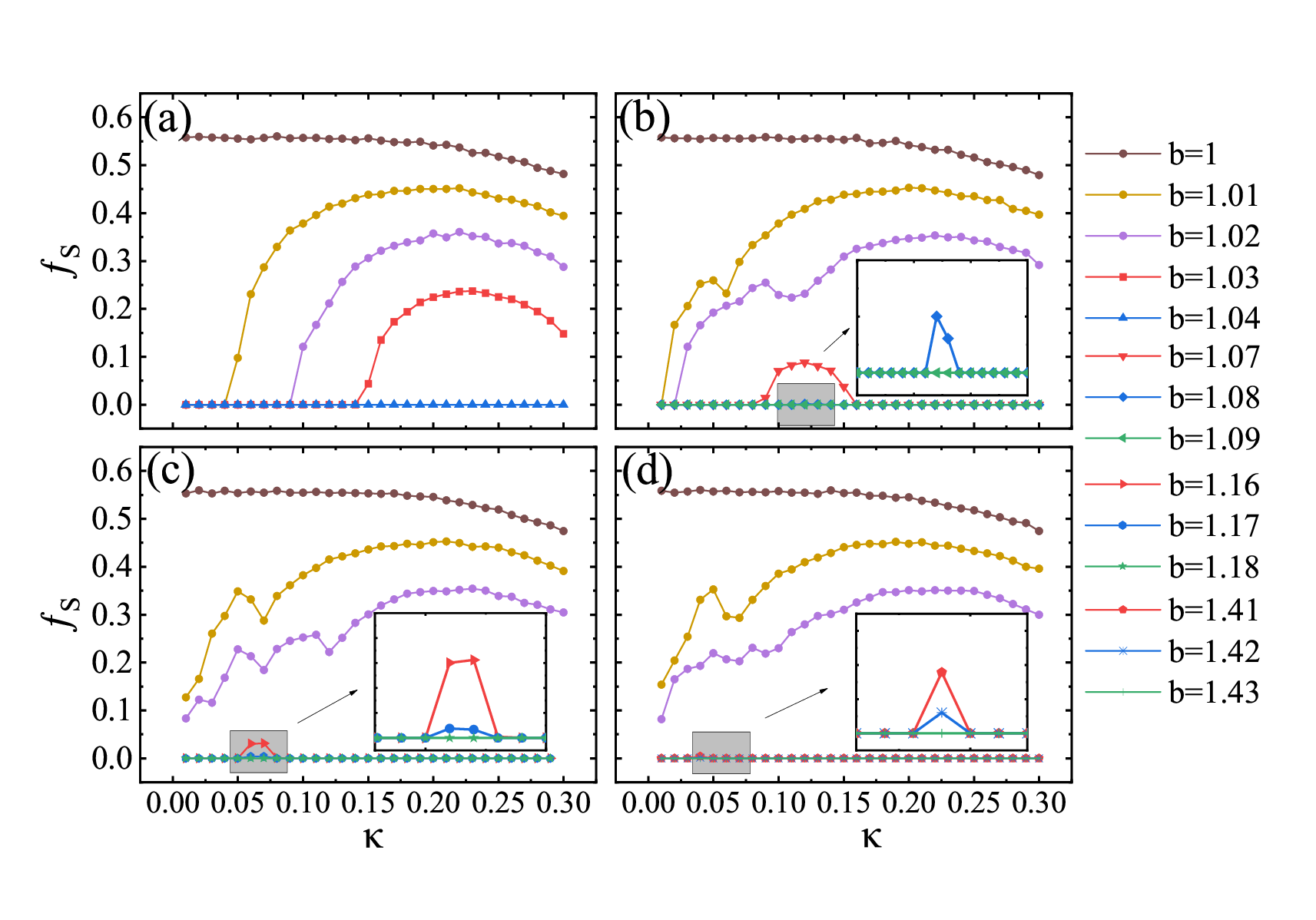}}
\caption{Diagrams of the average cooperation tendency over the population $f_S$ against the noise $\kappa$ for different defection temptations. (a) $m=1$, (b) $m=2$, (c) $m=4$, and (d) $m=8$.}\label{fig5}
\end{figure}

The effect of the noise $\kappa$ upon the average cooperation tendency over the whole population $f_S$ with different levels of cooperation tendency $m$ is also investigated and the result is shown in Fig.~\ref{fig5}. 
It shows much clearly that more cooperation tendency levels $m$ enlarges the region of average cooperation tendency $f_S$. Considering cooperators much preferring the cooperation tendency much close to the optimum one, the promotion of average cooperation tendency is the increase of the number of cooperators. 
When all cooperators are complete cooperators with $m=1$ shown in Fig.~\ref{fig5} (a), cooperators extinct 
at $b=1.04$ no matter how rational agents are. However, cooperators can survive even at $b=1.08$ when they have the half-cooperation tendency with $m=2$ in Fig.~\ref{fig5} (b). More levels of cooperation tendency $m$ allow cooperators to survive in a harder situation for $b=1.17$ with $m=4$ in Fig.~\ref{fig5} (c) and even for $b=1.42$ with $m=8$ in Fig.~\ref{fig5} (d). 
Furthermore, survival cooperators would like to be more rational when they are provided more cooperation tendencies. This is quite intuitive that agents need to be more rational to tell the difference between adjacent cooperation tendencies, which can be rather small when $m$ becomes large. 

\section{Conclusion} 
In summary, the effect of the strategy diversification, measured by the level of the cooperation tendency, on the evolution of the cooperation in the weak prisoners' dilemma is investigated. 
Because of the diversity of the strategy, agents can not only be complete cooperators or complete defectors but also also be partial cooperators and partial defectors at the same time. 
The evaluation of the cooperation can be measured by the fraction of cooperators, the ratio between the number of agents without the complete defective strategy, and the average cooperation tendency over the whole population due to the introduction of the cooperation tendency. 

The result of numerical experiments shows the cooperation level is significantly improved by introducing the cooperation tendency. 
Compared with only containing the complete cooperation and the complete defection strategy case, a higher cooperation diversity promises a higher cooperation level and brings the decline of the average cooperation tendency much slower and smoother when the temptation becomes higher. 
This is because that a higher cooperation diversity makes the cooperation efficiency, the average payoff per cooperation tendency, evolve slightly and form stable cooperator clusters. It also brings the optimum value of the average tendency of cooperators, defined by the ratio between the average cooperation tendency over all cooperators, at a certain defection temptation.
When the defection temptation is high enough, the cooperation vanishes because the cooperation efficiency cannot stay stable. 
Furthermore, the result above is independent of the noise and more rational agents can remain to be cooperators for a higher defection temptation.

\section*{Acknowledgements}
This research was funded by the National Natural Science Foundation of China under Grant Nos. $11947061$, by the Lanzhou Center for Theoretical Physics under Grant No. $12247101$.

\appendix

\printcredits

\end{document}